Low energy tracking and particles identification in the MUNU Time Projection Chamber at 1 bar. Possible application in low energy solar neutrino spectroscopy.

The MUNU collaboration: Z. Daraktchieva [a*], C. Amsler [b], M. Avenier [c], C. Broggini [d], J. Busto [a], C. Cerna [d], F. Juget [a], D.H. Koang [c], J. Lamblin [c], D. Lebrun [c], O. Link [b], G. Puglierin [d], A. Stutz [c], A. Tadsen [d], J.-L. Vuilleumier [a], J.-M. Vuilleumier [a], V. Zacek [e]

[a] Institut de physique, A.-L. Breguet 1, CH-2000 Neuchâtel, Switzerland
[b] Physik-Institut, Winterthurerstr. 190, CH-8057 Zurich, Switzerland
[c] Laboratoire de Physique Subatomique et de Cosmologie, IN2P3/CNRS-UJF, 53 Avenue des Martyrs, F-38026 Grenoble, France
[d] INFN, Via Marzolo 8, I-35131 Padova, Italy
[e] Universite de Montreal, C. P. 6128, Montreal, P.Q. Canada H3C 3J

**Abstract**

In this paper we present the results from the measurements made with the MUNU TPC at 1bar pressure of CF4 in the energy region below 1 MeV. Electron events down to 80 keV are successfully measured. The electron energy and direction are reconstructed for every contained single electron above 200 keV. As test the $^{137}$Cs photopeak is reconstructed by measuring both the energy and direction of the Compton electrons in the TPC.

---

[*] Corresponding author present address: Department of Physics and Astronomy, University College London, Gower Street, London WC1E 6BT, United Kingdom; *Email address*: zdar@hep.ucl.ac.uk  (Z. Daraktchieva)

## 1. Motivation

Time Projection Chambers (TPCs) have been used as tracking devices since the invention by Dave Nygren in the late 1970s. In the last years TPC's have been the main detector component in a large number of particle physics experiments due to their excellent particle recognition. In the low energy region gas TPC's are successfully used to study the neutrino properties [1] and to search for dark matter [2].

A gas TPC could be used to detect *pp* and $Be^7$ neutrinos and carry out solar neutrino spectroscopy. The neutrinos are detected via elastic scattering off electrons $v_e + e^- \rightarrow v_e + e^-$. The contribution of $v_\mu$ and $v_\tau$ neutrinos can be neglected because their cross section with electrons is about 1/7 of $v_e$ cross section. The initial electron direction can be determined from the tangent to the start of the electron track. This is possible if the tracks are long enough and if the spatial resolution is sufficient to distinguish between the beginning of the track from the end where the ionization is increased. This implies, for a given energy, that the gas pressure is below a given value. A reasonable goal is a threshold around 100 keV on the electron [3]. The neutrino energy can be reconstructed from the electron energy and direction thereafter.

MUNU, a time projection chamber installed at the Bugey reactor to study antineutrino electron scattering $\bar{v}_e + e^- \rightarrow \bar{v}_e + e^-$, was used to demonstrate the feasibility of such a scheme. The results of these measurements are presented here. Particle identification properties of gas TPC's, which should lead to a good event selection and a low background rate, will also be addressed.

## 2. Experiment

The essential features of the MUNU detector have been presented previously [4, 5]. Briefly the detector consists of a $1m^3$ Time Projection Chamber (TPC) housed in an acrylic vessel (90 cm diameter and 162 cm long) filled with 3.8 kg of pure CF4 gas (1.06 x $10^{21}$ electrons per $cm^3$ at STP). The gas was selected because of its high electron density, good drifting properties, low Z (which reduces multiple scattering) and its absence of free protons. Most data were taken at a pressure of 3 bars. The data reported in this paper have been taken at a pressure of 1 bar with an energy threshold of 100 keV for the recoiling electrons. As shown in Fig.1 at one end of the TPC, a cathode is mounted and was held at a negative high voltage. It defines the drift field along with field shaping rings outside the acrylic vessel and a grid at the other end. The cathode was held at a potential of *–17 kV* for the 1 bar measurements, while the grid was at *–1.28 kV*. This led to a drift velocity of *2.14 cm.µs$^{-1}$*. An anode plane with *20 µm* diameter wires and a pitch of *4.95 mm*, separated by *100 µm* potential wires, is placed behind the grid to amplify the ionization charge. The integrated anode signal gives the total energy deposit. A pick-up plane with perpendicular x and y strips (pitch *3.5 mm*) provides the spatial information in the x-y plane perpendicular to the TPC axis (z-axis). The spatial information along the z-axis is reconstructed from the time evolution of the signal. The acrylic vessel TPC is immersed in a steel tank (2m diameter and 3.8 m long) filled with 10 m$^3$ of an organic liquid scintillator (NE-235). The liquid scintillator is viewed by 48 photomultipliers and serves as an anti-Compton detector and a veto for the cosmic muons. The liquid scintillator and steel vessel both serve as a low activity active shielding. In addition, there is 8 cm thick borated polyethylene and 15 cm thick lead shielding to absorb the external neutrons and γ rays.

**3. Energy calibration of the TPC at 1-bar**

For the energy calibration of the TPC at 1 bar pressure of CF4, $^{54}$Mn and $^{137}$Cs γ radioactive sources were used with gamma energies of 834 keV and 662 keV respectively. These sources were introduced into the anti Compton detector through an acrylic pipe at an axial distance as low as 26.4 cm from the centre of the detector (1 cm from the TPC side wall) on the anode side. The radial position was varied. Most detected photons undergo only one Compton scattering in the active volume of the TPC before leaving it. The calculated energy of the Compton edge is found to be 638 keV and 478 keV for $^{54}$Mn and $^{137}$Cs respectively. However the Compton edge in the TPC is smeared with the resolution of the detector which complicates the direct calibration. We performed Monte Carlo simulations with the GEANT 3 code. The Compton electron spectrum was calculated in the TPC for various resolutions of the detector. The Gaussian width of the detector response was obtained by comparing with the measured spectrum [6]. Fig. 2 shows the measured spectrum for $^{137}$Cs along with the best simulation with *σ/E=6.8 %*.

**4. Energy resolution of the TPC at 1-bar**

The energy resolution of a gas TPC depends on one hand on the electrical property of the gas itself (ionisation of the gas, attachment and multiplication) and on the other hand on the quality of the detector (homogeneity of the gain on the anode plane and charge attenuation). At 1-bar pressure the effective attachment of CF4 is found to be 85 % [7]. That means that 15 % of the primary ionisation electrons will survive and reach the multiplication phase of the avalanche. This factor is about 4 times better than that at 3 bars (4%). The energy resolution of the TPC was measured as described above with $^{137}$Cs and $^{54}$Mn radioactive sources and was found to be 6.8 % (1σ) and 6.1 % (1σ) at 478 keV and 638 keV respectively. It scales with energy with the following empirical $E^{0.57}$ law. Thus the resolution varies from 10 % at 200 keV to 5 % at 1 MeV, being about 2 times better than the resolution at 3-bar [5].

**5. Particle identification**

Low background is essential in a solar neutrino experiment. The detector must be made from highly pure materials. In addition, the detector must use an event signature as selective as possible. Particle identification is crucial in this context.

## 5.1 Compton electrons

A Compton electron from a gamma activity which originates from inside the TPC could mimics an electron from neutrino scattering if the gamma does not reinteract in the active volume of the TPC. In MUNU however the escaping photon will be detected in the veto counter with high efficiency The anti-Compton trigger was set at the energy threshold of 100 keV and used a multiplicity criteria seeking at least 5 phototubes hits. The suppression efficiency for Compton electrons from γ background is found to be 97 %.

## 5.2 Muons

With an overburden equivalent to 20 meter of water the MUNU detector is exposed to a constant flux of cosmic muons of 32 $m^{-2}$ $s^{-1}$. The deposited mean energy loss per cm in the CF4 at 1-bar pressure is 6.4 keV.$cm^{-1}$ as estimated from a visual scan [8] of simulated muons. A cosmic muon has to cross more than 1 m of liquid scintillator if it hits TPC, leaving behind large prompt signal in the phototube. This signal of the anti- Compton detector is used to veto the muon associated events with almost 99 % efficiency.
In Fig.3 are displayed the *xz*, *yz* and *xy* projections of a cosmic muon track in the TPC.

## 5.3. Alpha particles

The alpha particles from the natural radioactivity deposit all of their energy in a very short range in CF4. For example the track length of a 5 MeV α particle, as it shown in Fig.4, is 28 mm in 1-bar of CF4.

### 5.4. e⁺e⁻ pairs

The e⁺e⁻ pairs from the cosmic and natural radioactivity can clearly be seen in the TPC and identified. The example in Fig.5 shows two tracks in 1bar of CF4.

### 5.5. Rare events

TPC can also detect some rare events, for example protons. A proton can be identified from the high ionization and short straight track as it is shown in Fig.6.

### 5.6. Contained single electron events

Data were accumulated during 5.2 days live time reactor on as reported in ref. [9]. Electron events down to 70 keV have been measured in the TPC. Examples of low energy single electrons of 80 keV and 540 keV, respectively are displayed in Fig.7a and Fig.7.b. The low ionizing start of the track and the increased energy deposition at the end can be seen in both projections. The counting rate of single electrons (neutrino trigger) above 100 keV is 0.4 s$^{-1}$: 0.05 s$^{-1}$ from cosmic muons and 0.31 s$^{-1}$ from Compton electrons and natural radioactivity. The deadtime, about 50 % of the total time, is mostly due to the low energy threshold and anti-Compton veto, leading to a high data readout and transfer time.

Above 150 keV the electron tracks are long enough and can be reconstructed with good precision. Here the neutrino candidates are carefully selected being single electrons fully contained in a 42 cm TPC radius, with no energy deposition above 100 keV in the anti-Compton detector in the preceding 80 μs corresponding to the longest possible drift time in the TPC. In the data analysis an additional software energy filter of 200 keV is applied in order to suppress some point like electrons. For single electrons fully contained inside the TPC above 200 keV the rate is 2900 counts per day (cpd) live time, i.e. 760 cpd/kg. The containment efficiency was computed by GEANT 3 simulations, 10000 single electrons were generated randomly inside the TPC with flat energy distribution between 20 keV and 2 MeV. The containment efficiency of the TPC at 1 bar then is found to vary from 85 % at 200 keV, 50 % at 400 keV to 10 % at 800 keV. The energy distribution of measured contained electrons is in good agreement with the results from simulations (Fig. 8). The neutrino energy $E_\nu$ is reconstructed from the scattering angle with respect of reactor-detector axis $\theta_{rea}$ and the electron recoil energy $T_e$ measured in the TPC. The angular resolution, obtained from the visual scan of simulated electron tracks, is found to vary from 15 ° (1σ) at 200 keV to 10° at 600 keV.

Electrons from neutrino scattering are emitted forward with regards to the reactor core, while the background from activity is isotropic. We used the kinematical procedure of ref. [9] to determine the signal plus background rate in the forward direction and the background rate in the backward direction. After applying the fiducial, geometrical and kinematical cuts we have measured the forward minus backward rate corresponding to the signal from antineutrino electron interaction of 2.89 ± 2.39 counts per day. This is consistent with the expected number assuming no neutrino magnetic moment, 0.49 ± 0.12 cpd. Due to the limited statistics and uncertainties in the low neutrino energy spectrum this result is not used to study the neutrino magnetic moment. However it proves that a gas TPC can be used to measure the energy and directions of the recoil electrons at least down to 200 keV.

## 6. Reconstruction of Cs photopeak

To investigate more precisely the reconstruction of the neutrino energy from the electron scattering angle and kinetic energy, the TPC was exposed to gammas from a $^{137}$Cs source (662 keV). These measurements were performed at a pressure of 1 bar of CF4.

The γ-ray from $^{54}$Mn at 835 keV at 3bar pressure of CF4 was measured by MUNU previously [5].

The Compton scattered photon is measured in the scintillator. The recoil electron track and energy are measured in the TPC. The Compton electrons initial direction is obtained by a visual fit. The angle $\theta_{source}$ with respect to the source axis (perpendicular to the detector-reactor axis) is calculated from this fit. The initial photon energy $E_\gamma$ is reconstructed thereafter from the scattering angle $\theta_{source}$ and the electron recoil energy $T_e$ measured in the TPC.

The reconstructed 662 keV photopeak from the $^{137}$Cs source at 1 bar of CF4 together with the cosine of the scattering angle $\theta_{source}$ and electron recoil energy from Compton scattering are presented and compared with Monte Carlo simulations in Fig.9. The energy resolution from the Compton edge at 1 σ is 32.6 keV at 480 keV. The angular resolution of the Compton recoil spectrum above 150 keV is $\sigma_\theta = 11.6°\pm 0.9$. This is in good agreement with the above mentioned determination from Monte Carlo simulations. The width of the reconstructed energy peak at 1σ is 220 keV at 662 keV.

Monte Carlo simulations have been performed in order to check the feasibility of using a gas TPC to detect neutrinos from the Sun [3]. More in details the reasonable volume of a TPC should be about 1000 $m^3$ with 3.7 *ton* of CF4. The counting rate is expecting to be 5 solar event/day from both *pp* and $^7$*Be* neutrinos at a threshold of 100 keV of electron recoil energy. The simulations have demonstrated that after 2 years of data taking with such a TPC operating at 1 bar pressure, it is possible to reconstruct the energy spectrum of the *pp* and $^7$*Be* neutrinos via elastic scattering with electrons. The σ of the reconstructed $^7$*Be* peak is found to be 230 keV at 862 keV, which result is in very good agreement with the result of the reconstructed here $^{137}$*Cs* photo peak at 1-bar. It also demonstrates that the MUNU technology could be used in the future for solar neutrino spectroscopy if instead of photons one chooses the low energy neutrinos from the sun.

## 7. Conclusion

The measurements made with the MUNU TPC at 1 bar of CF4 are presented. The background is well understood and separated from the good single electron events. The TPC can work successfully at energies as low as 100 keV. The single electron events down to 80 keV (4.5 cm length) are measured in the TPC. Above 200 keV the electron tracks have been analysed, the direction and energy are reconstructed.

Also here we presented the reconstructed 662 keV $^{137}$Cs full γ energy from Compton scattering for 1 bar of CF4. All these results are encouraging and demonstrate the possibilities of future low energy solar neutrino spectroscopy with a gas TPC, installed in a very low background environment underground facility.

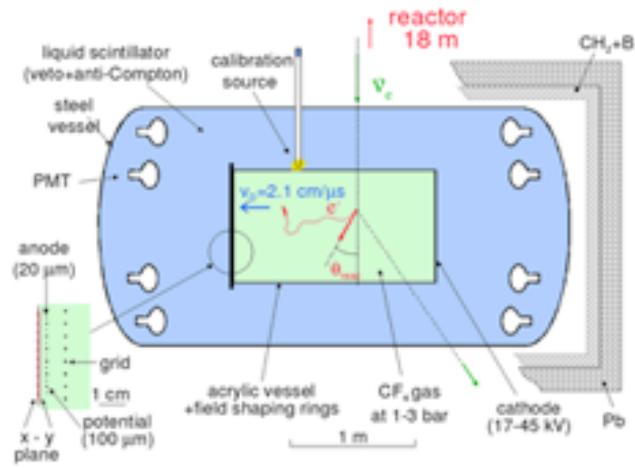

Fig.1 The MUNU detector

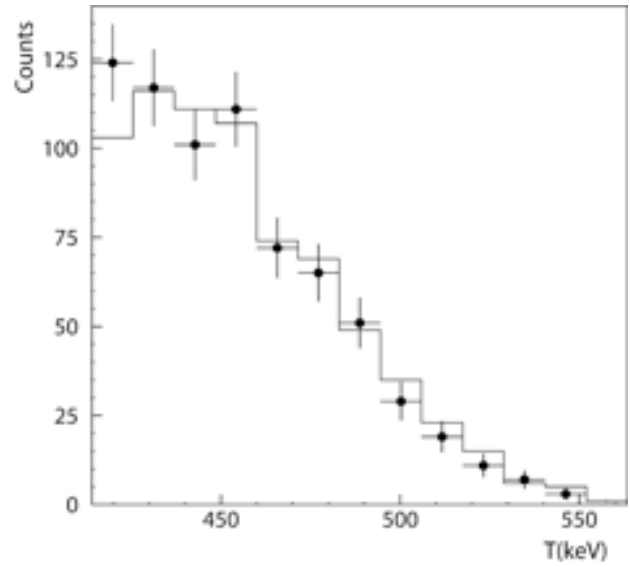

Fig.2. A Compton spectrum of $^{137}$Cs in the TPC compared with Monte Carlo simulation (filled circles).

a.  b.

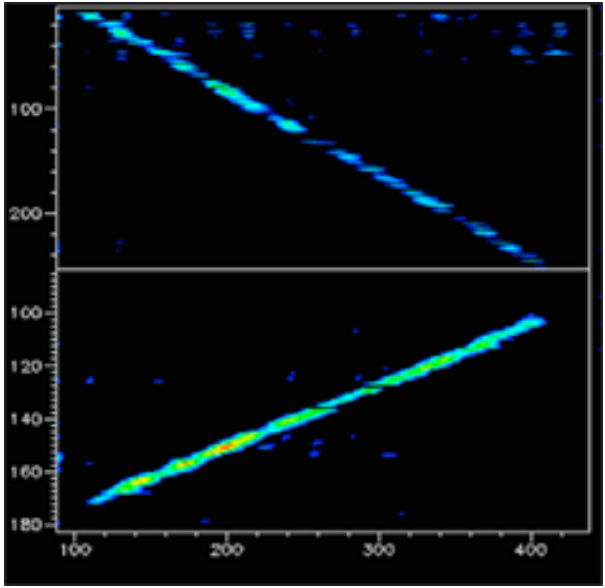 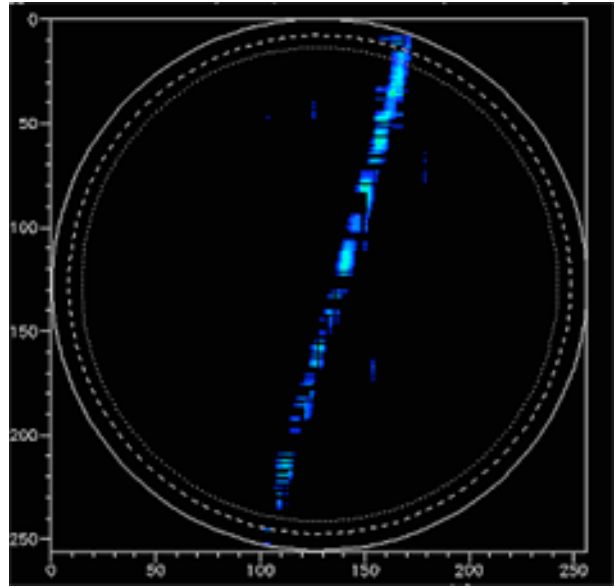

Fig.3 A cosmic muon event in 1-bar of CF4: (a) *xz* projection (top), *yz* projection (bottom), and (b) *xy* projection. The binning is 3.5 mm for *x* and *y* and 1.7 *mm* (80 *ns* for *z*).

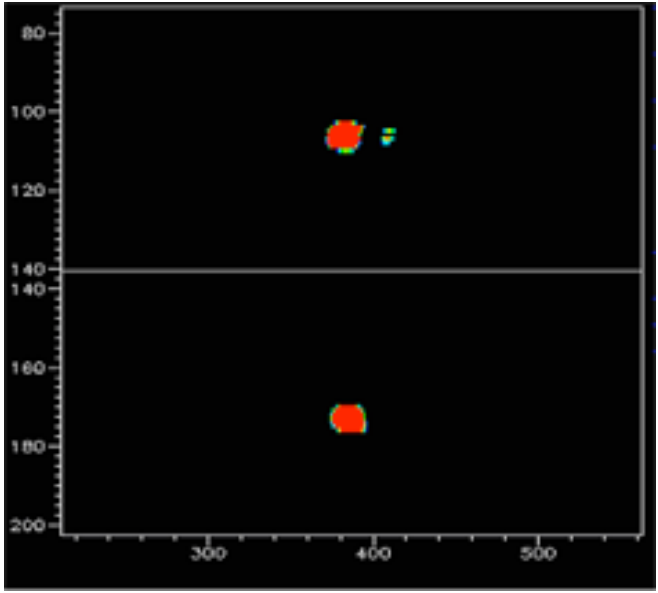 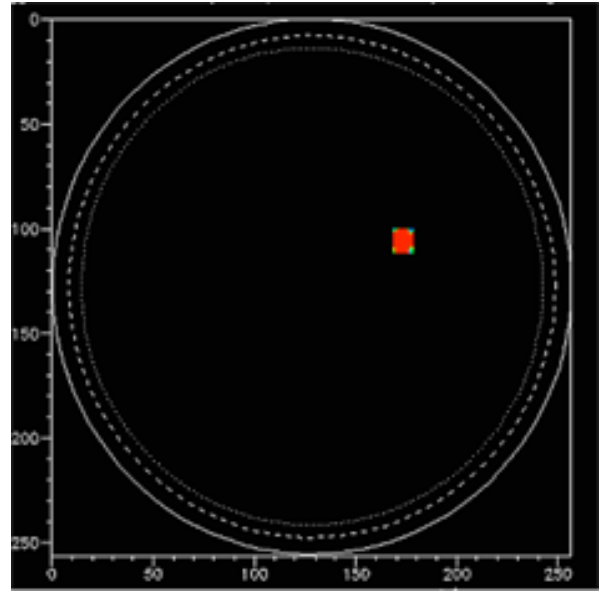

Fig.4. An alpha particle event: (a) *xz* (top), *yz* (bottom) and (b) *xy*.

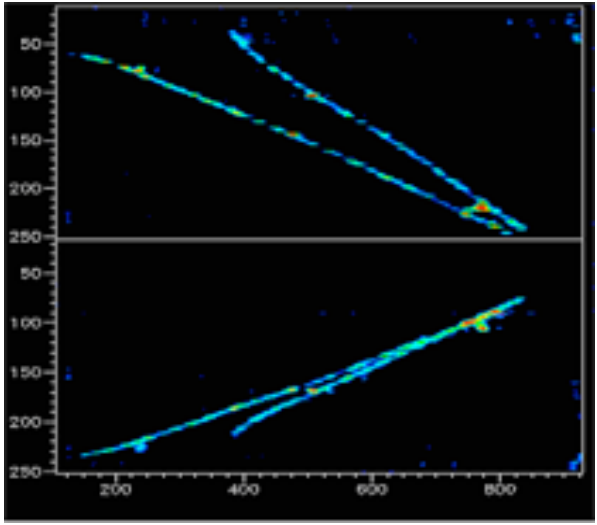 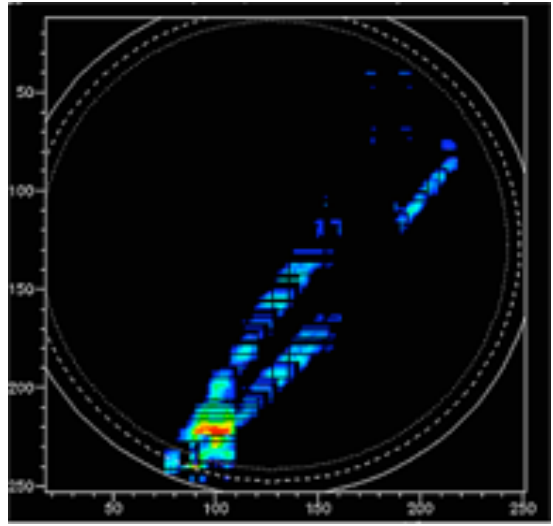

Fig.5. Two e⁺e⁻ pairs: (a) *xz* (top), *yz* (bottom) and (b) *xy*.

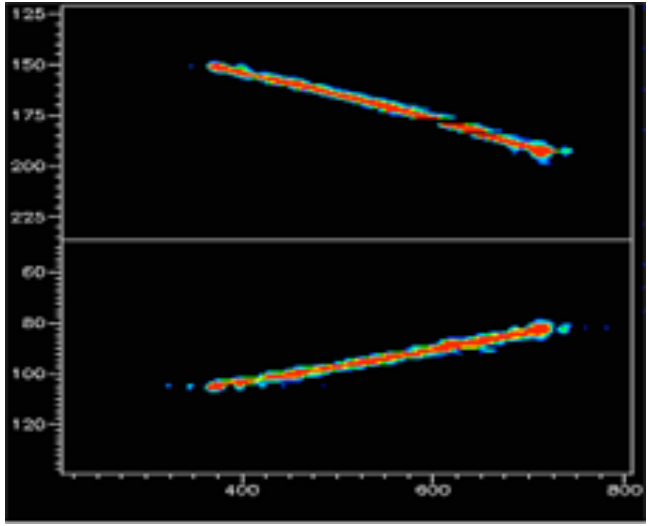 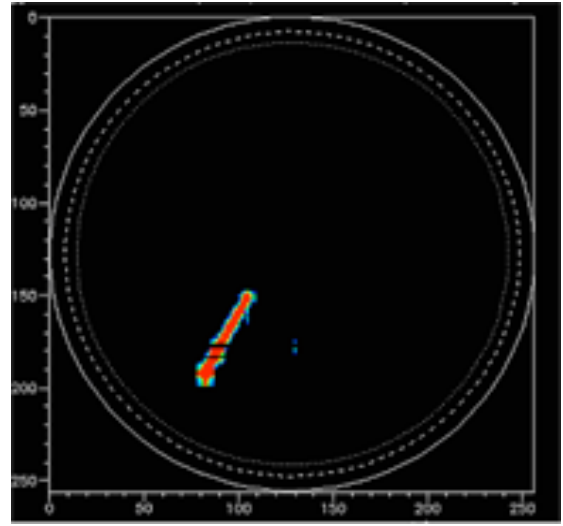

Fig.6. A proton: (a) *xz* (top), *yz* (bottom) and (b) *xy*.

a. 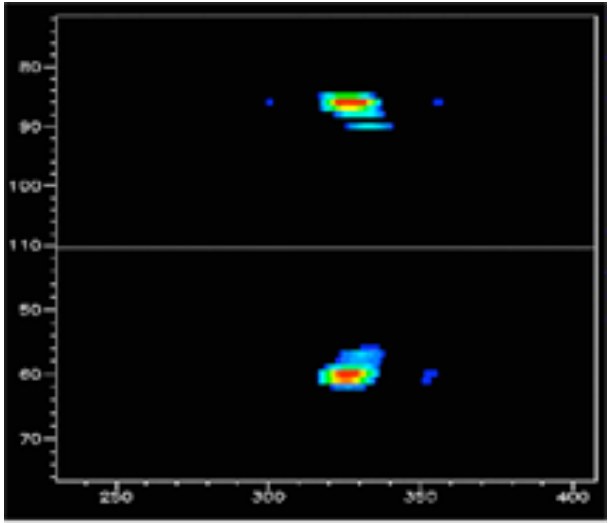 b. 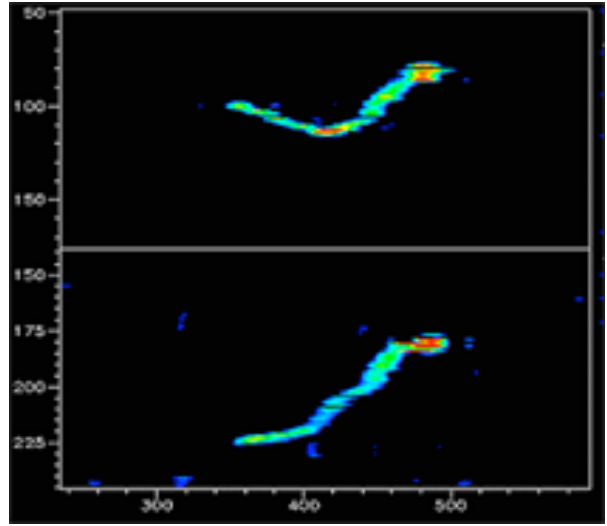

Fig.7. Low energy single electrons of (a) 80 keV (4.5 cm ) and (b) 540 keV (35 cm) at 1 bar in the TPC: *xz* (tops) and *yz* (bottoms).

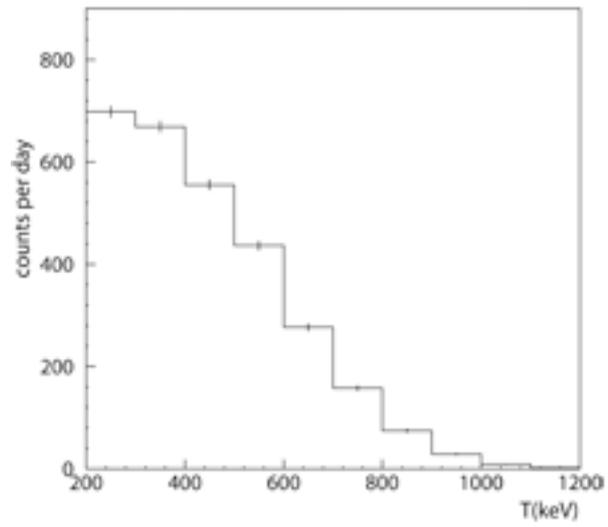

Fig. 8. Contained single electron events above 200 keV.

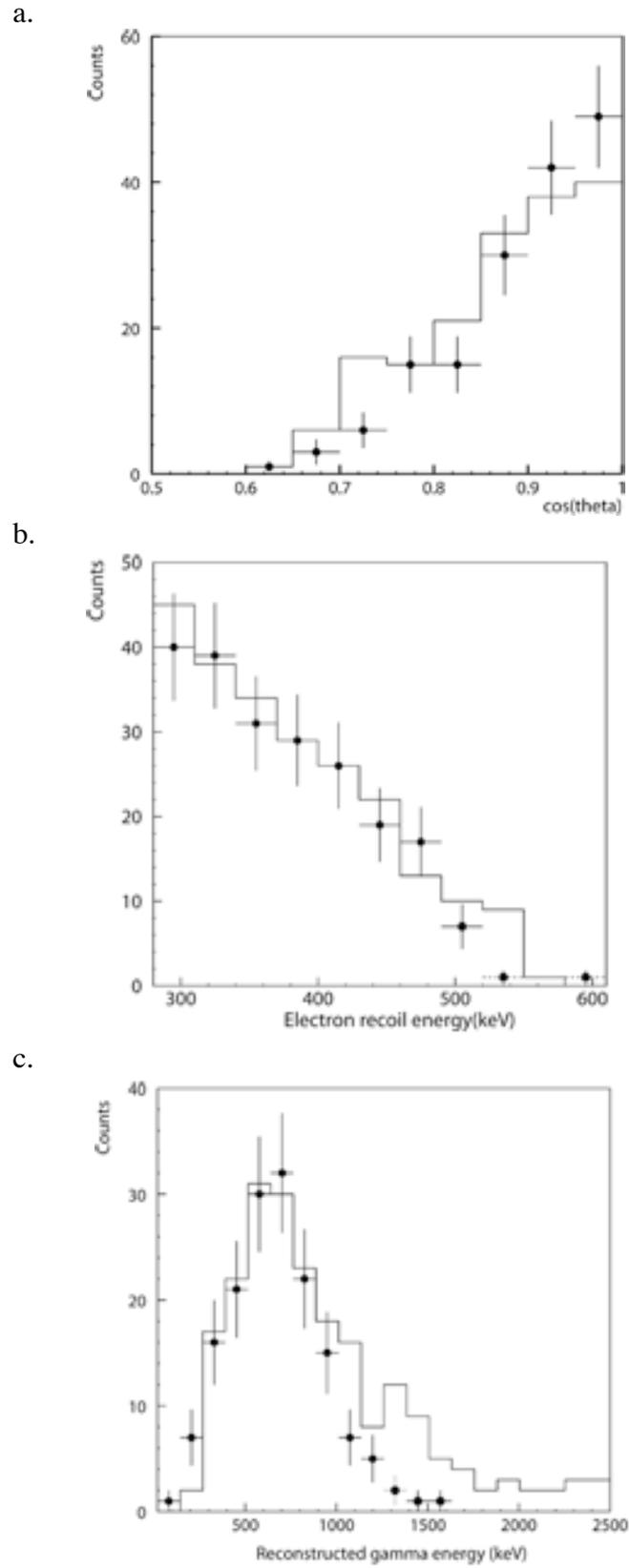

Fig.9. (a) Scattering angle $\theta_{source}$, (b) electron recoil energy $T_e$ and (c) reconstructed gamma energy $E_\gamma$ (c) for $^{137}$Cs, compared with Monte Carlo simulations (filled circles).